# A Two-Layer Component-Based Allocation for Embedded Systems with GPUs

**Gabriel Campeanu [1],*** and **Mehrdad Saadatmand [2]** 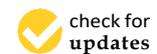

[1] Bombardier Transportation, 722 14 Västerås, Sweden
[2] RISE Research Institutes of Sweden, RISE SICS Västerås, 722 12 Västerås, Sweden; mehrdad.saadatmand@ri.se
* Correspondence: gabriel.campeanu@rail.bombardier.com

**Abstract:** Component-based development is a software engineering paradigm that can facilitate the construction of embedded systems and tackle its complexities. The modern embedded systems have more and more demanding requirements. One way to cope with such a versatile and growing set of requirements is to employ heterogeneous processing power, i.e., CPU–GPU architectures. The new CPU–GPU embedded boards deliver an increased performance but also introduce additional complexity and challenges. In this work, we address the component-to-hardware allocation for CPU–GPU embedded systems. The allocation for such systems is much complex due to the increased amount of GPU-related information. For example, while in traditional embedded systems the allocation mechanism may consider only the CPU memory usage of components to find an appropriate allocation scheme, in heterogeneous systems, the GPU memory usage needs also to be taken into account in the allocation process. This paper aims at decreasing the component-to-hardware allocation complexity by introducing a two-layer component-based architecture for heterogeneous embedded systems. The detailed CPU–GPU information of the system is abstracted at a high-layer by compacting connected components into single units that behave as regular components. The allocator, based on the compacted information received from the high-level layer, computes, with a decreased complexity, feasible allocation schemes. In the last part of the paper, the two-layer allocation method is evaluated using an existing embedded system demonstrator; namely, an underwater robot.

## 1. Introduction

Nowadays, embedded systems become more and more common in the daily life. Modern embedded systems, characterized by new and demanding functionalities, deal with huge amount of information resulted from the interaction with the environment. For instance, the Google autonomous car ( i.e., the Waymo project) handles 750 MB of data per second that is produced by its sensors (e.g., LIDAR). The huge amount of information needs to be processed with a particular performance, in order to satisfy the system requirements. For example, the autonomous Google car needs to process its captured data in real-time in order to detect various objects and pedestrians, to avoid accidents. One solution to enhance the processing capacity of embedded systems comes from the usage of embedded boards with Graphics Processing Units (GPUs). A GPU is a processing unit that is equipped with hundreds of computation threads, excelling in parallel data-processing.

Although, on one side, the use of GPU increases the system (parallel-processing) performance, on the other side it increases the complexity of the system design. In particular, the software-to-hardware allocation is already not an easy task: when having several processing units of different kinds and with different capabilities, a major design challenge will then be in finding an optimal allocation of software artifacts (e.g., components) onto the processing units in a way that system constraints are also met and not violated. For allocating a set of $n$ software artifacts onto $m$ processing units, a total number of $m^n$ combinations are to be considered [1]. The challenge, as mentioned, is then to find, from all combinations, a single permutation as an optimal allocation scheme with respect to the constraints and characteristics of the system. With the GPU in the landscape, the allocation becomes even more complicated and challenging. The software is characterized now, besides the properties regarding the CPU resources, with properties that refer to GPU such as the GPU memory usage or the execution performance on the GPU. Similarly, the platform also has characteristics regarding the CPU but also the GPU hardware. Hence, the allocation challenge is increased due to the extra application properties and hardware characteristics that must be considered. In short, the challenge of finding optimal allocations of software artifacts to hardware has increasingly attracted attention, especially with the advent and growing prevalence of heterogeneous hardware platforms and increasing use of software in mission-critical applications. In [2], for instance, Baruah has discussed the challenge of allocating a set of recurring tasks in real-time systems onto processing units of different kinds while respecting all timing constraints, and has identified this to be an NP-hard problem. One way to relieve the allocation challenge and cope with its complexity is by managing the amount of information that is fed to the allocator to make allocation decisions. This is the main topic and solution that we introduce and investigate in this paper. In other words, in this current work we do not focus mainly on how to derive and what will be an optimal allocation scheme for a system (which is the main focus in [1,3]), but rather how the burden on the allocator can be relieved to relax the complexity of the allocation process in general.

In this work, we use the component-based development (CBD) to construct embedded systems with GPUs. In general, this software engineering methodology promotes the construction of applications by composing existing software units called software components. CBD is used and promoted in industry to construct embedded systems; such as in AUTOSAR [4] which is now the de-facto standard in automotive industry, and IEC 61131 [5] used to develop programmable logic controllers (PLCs). In the context of component-based embedded systems with GPUs, we focus on the component-to-hardware allocation, proposing a semi-automatic allocation method. When using platforms with GPUs, the allocation challenge increases even more due to the (higher) complexity of the software and hardware. The software in such systems is composed of: (i) (traditional) components that have requirements on common resources (e.g., CPU load, RAM memory usage), and (ii) (GPU-specialized) components that have GPU requirements (e.g., number of GPU threads usage). These GPU-specialized components, although they contain small CPU functionality (e.g., activities to trigger execution on GPU), are seen as components with only GPU computation. Having a pool of (CPU and GPU-based) components, many alternatives with the same functionality may result. For example, a vision system may have two alternatives, where one alternative contains only components with CPU functionally, while the other alternative contains only components with GPU computations. Regarding the hardware, the platform encloses, besides the traditional CPU, the GPU that has different characteristics such as the available GPU memory.

The aim of our work is to alleviate the allocation challenge by mitigating the increased amount of (software and hardware) information. In the context of applications with multiple alternatives, and CPU–GPU hardware, the allocator does not need to take in consideration all the system information. For instance, the information that describes the component communication from inside the alternatives may be neglected. An implicit constraint considered in this work is, due to the closely connected nature of the GPU to the CPU, the allocator needs to deploy a (entire) variant that has GPU computations, onto a CPU–GPU processing node. Enforcing this requirement improves the overall system performance.

As a solution, we propose a two-layer architecture to decrease the allocation effort. Both layers describe a same system that has GPU computation; the difference resides in the level of information that characterize each layer. The first layer, seen as regular description of the architecture, encloses all information (e.g., component communication links) of all alternatives. The second layer compacts different alternatives with the same functionality into single components with multiple variants. Each of the variants of the resulted components, is characterized by a set of properties that reflect the requirements of all components contained by its corresponding alternative.

To abstract certain information, the allocator uses the first layer description and selects a suitable component alternative. Once the allocation scheme is computed, the selected alternative is unfolded with the corresponding structure from the second layer. The core idea of a two-layer allocation method is to decrease the information load and constraints that may increase the overhead of the allocator. Another benefit of our solution is the increased scalability characteristic, where our allocator may handle more complex systems (e.g., characterized by a high number of components) due to the decrease of the information and constraints load.

In this work, we use in the evaluation section an already existing (constraint-based) allocator that we constructed in a previous work [3]. Our approach is independent of what allocator is used or how it is implemented. By proposing the two-layer allocation design, we improve the scalability of the allocator used for heterogeneous CPU–GPU systems, and decrease its burden.

The remaining of the paper is organized as follows. Section 2 introduces the context details of this work. Furthermore, the section presents a running example, and, based on it, a way to develop component-based systems for heterogeneous embedded systems. The overview of our method is described in Section 3 while Section 4 evaluates our method using the introduced running example. Related work is covered by Section 5 followed by conclusions and future work in Section 6.

## 2. Component-Based Design for Heterogeneous CPU–GPU Architectures

The growing complexity and size of embedded systems emphasizes the use of appropriate development methods that can cope with these issues and scale well. Component-Based Development (CBD) is a promising approach in this regard which promotes building a system out of already existing components, as opposed to building it from scratch. In other words, CBD enables *reusability* in software development by building a system as an assembly of components [6,7] selected from a repository of verified existing components. Considering the constraints of embedded systems in terms of available resources, how software components are allocated onto the hardware platform can play an important role in the performance of the system and optimal use of the resources. This is, however, not a trivial task and as the number of software components as well as processing units and computing nodes on the hardware platform increases, different combinatorial allocation schemes need to be evaluated in order to determine an appropriate one. For this reason, having an automated solution for evaluation of allocation schemes is necessary. One aspect that adds to the challenge of allocating software components onto hardware platforms is the move towards the use of heterogeneous hardware architectures such as multi-core CPU and GPU. Use of GPUs along with CPUs is particularly interesting as it can provide increased computation power and diversity due to the parallel processing capability of GPUs. This brings along additional constraints that need to be taken into account for allocation of software components to hardware. For instance, a GPU cannot be used independently of a CPU, and it is the CPU that triggers all GPU specific operations such as data transfer between the main memory unit (i.e., RAM) and the GPU memory system. Therefore, there is constant and high communication between these two processing units.

From the perspective of the processing unit, software components can be categorized as: (i) those that require only CPU for their functionality (ii) components that use GPU (in addition to CPU) to fulfill their functionality. For the rest of the paper, these types of components will be referred to as GPU components. A specific functionality may be implemented as any of these component types. Therefore, in the component repository both types of components may exist as different implementation versions of a specific function. For instance, for an image processing component we may have three different implementation versions in the repository; one that uses CPU only (type i), and two others as GPU-based implementations that require GPU as well (type ii). Each of these three versions can have different properties and characteristics in terms of resource usage and utilization. For example, the two GPU-based component implementations can have different resource usage properties with respect to GPU memory and GPU computation threads.

In the next paragraphs, we introduce a case study that will be used to present our solution. The case study is an underwater robot [8] that autonomously navigates under water and executes various missions such as finding buoys. The robot is a typical embedded system that contains sensors (e.g., cameras), an embedded board with an incorporated GPU, and actuators (e.g., thrusters). An interesting part of the robot is the vision part which is described in Figure 1. The vision (sub-)system is developed with a state-of-practice component model i.e., the Rubus component model [9]. This particular component model follows a *pipe-and-filter* interaction style, where each component computes its received data and sends it to the next connected component(s). The vision system consists of six components as follows. The first two components (i.e., *Camera1* and *Camera2*) receive raw data from two camera sensors, convert it into readable color (i.e., RGB) frames and forward it to the *MergeAndEnhance* component. After the two frames are merged into a single RGB frame and its noise is reduced, the *ConvertGrayscale* component converts it into a grayscale format and sends it to the *EdgeDetection* component. This component converts the frame into a black-and-white frame, where the white lines delimit the objects from the frame. Finally, the *ObjectDetection* component detects a particular object, such a buoys.

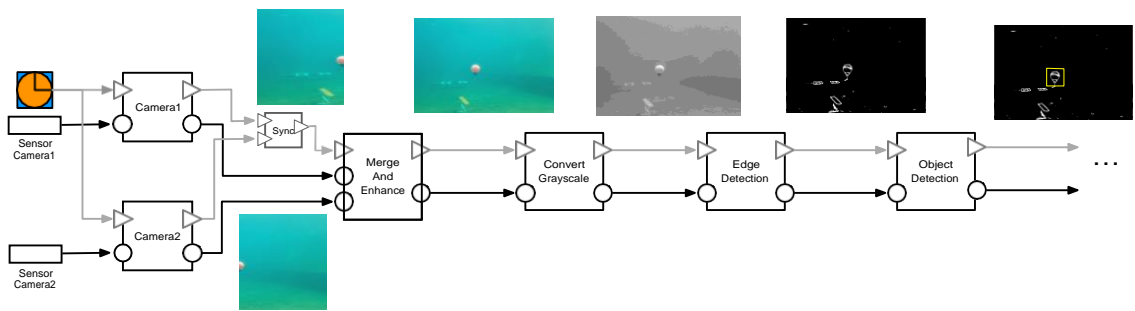

**Figure 1.** The vision system of an underwater robot.

Due to the fact that several components (i.e., *MergeAndEnhance*, *ConvertGrayscale*, *EdgeDetection* and *ObjectDetection*) have functionalities (i.e., image processing) that can be executed on the GPU, the vision system may have different alternatives. Figure 2 presents possible alternatives of the vision system. Assuming that we have a repository with ten components where there are components with the same behavior but constructed to be executed on different processing units (i.e., either on CPU or GPU). For instance, there is the *EdgeDetection(GPU)* component that is constructed to be executed on the GPU and there is *EdgeDetection(CPU)* that has the same behaviour but it requires to be executed on the CPU. The total amount of alternatives that can be constructed by using the repository components is six, as illustrated in the figure. The first alternative, where all components are executed on the CPU, has a low performance but also zero-demand on the GPU. This alternative can be selected to be used in a system that is not equipped with a GPU or in a system that possesses a GPU but it is used by a different part of the system. The last alternative, containing four components that need to be executed on the GPU, has the highest performance compared to the rest of the alternatives but it also has high GPU requirements (e.g., GPU memory and computation threads usage). The other four possible alternatives contain variations of components with different requirements on the CPU and GPU.

In this context where there are several alternatives and each one contains different components versions characterized by distinct characteristics, the information load on the allocator is much, influencing, in a negative way, the allocation efficiency. Furthermore, because of the tight connection between CPU and GPU, the components executed on a GPU are desired to be placed, alongside with their connected CPU-based components, on the same CPU–GPU chip. In the opposite case, placing a GPU-component on a different CPU–GPU chip than its connected CPU-component brings additional communication overhead which negativly influences the system performance. In general, the allocation complexity is directly influenced by the number of considered alternatives and by the number of the components and their versions included in the alternatives.

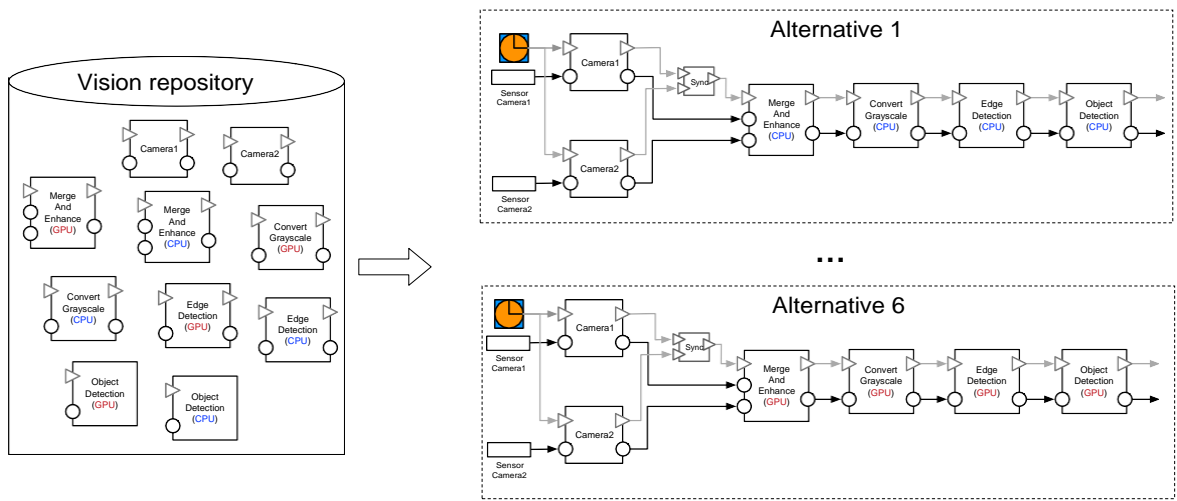

**Figure 2.** The vision system alternatives. GPU, graphics processing units; CPU, central processing unit.

## 3. Solution Overview

To diminish the burden of the allocation process, we introduce a two-layer allocation method. The layers correspond to a two-layer architecture view of the system, where the bottom layer describes a detailed system (i.e., composition of different component versions), while the top layer abstracts the complexity of the detailed system, by compacting the resulting alternatives into units (i.e., multi-variant units) that behave as regular components. The resulted components are characterized by different alternatives, where an alternative contains the properties that correspond to all its enclosed components.

Figure 3 depicts the vision system alternatives compacted into a multi-variant component where each alternative is characterized by set of properties. These properties are derived from the properties of the enclosed components. For instance, the alternative 6 where there are four GPU-based components, is characterized by a GPU memory property that describes the GPU memory usage of the enclosed components. This property is derived by e.g., summing the GPU memory usage of all four GPU-based components. As the GPU threads are highly reusable between GPU-based components, the alternative property that describes the number of GPU threads usage is the highest value of threads usage among the four GPU-based components. Furthermore, some attributes may be abstracted away. For example, due to the fact that connected (CPU- and GPU-based) components are enclosed together into single units and implicitly they will be allocated on the same CPU–GPU chip, the bandwidth property required for data transfer between two connected components, is abstracted away.

Figure 4 illustrates the overview description of our solution, that contains six stages as follows:

1. The first stage refers to the component pool from which the system developer constructs the application. The components from the repository may be provided by a 3rd-party or developed in-house. The repository contains regular (CPU-based) components but also components with GPU capability. For instance, there are two component versions (i.e., *C2 GPU* and *C2 CPU*) with the same behaviour but different (hardware) requirements.

2. Using the available components from the repository, the system architect composes them in different alternative systems, as described in the second stage. For example, while the first alternative uses *C1 GPU*, the second system alternative contains *C1 CPU*. All the system alternatives have the same behavior and different requirements. For example, while the first alternative requires GPU memory for two components, the second alternative has only one component with GPU requirement.
3. The third stage compacts all the component variants into multi-variant components. For example, the several alternatives that contain three components are grouped into a unit with different alternatives. These alternatives have the same functionality but with different (CPU and GPU) requirements. This is a simplified system architecture in which we abstracted away some information.
4. The information from the previous stage is forwarded to an allocator which we assume it already exists. Our approach is independent of how the allocator itself is implemented and based on which solver. Other information is fed to the allocator such as the description of the platform or different system constraints. In this stage, the allocator computes component-to-hardware allocation schemes where each component is mapped to a single processing unit.
5. The fifth stage describes the system architecture based on the result computed by the allocator. The architecture contains only (single-variant) components, that is each component has a single set of requirements and is allocated to a particular processing unit.
6. The last stage contains a fully detailed system architecture where the (single-variant) components from the previous stage are unfolded (when it is possible). The figure illustrates the detailed system architecture where the alternative selected by the allocator is unfolded into three connected components, i.e., *C1 GPU*, *C2 GPU* and *C3 CPU*.

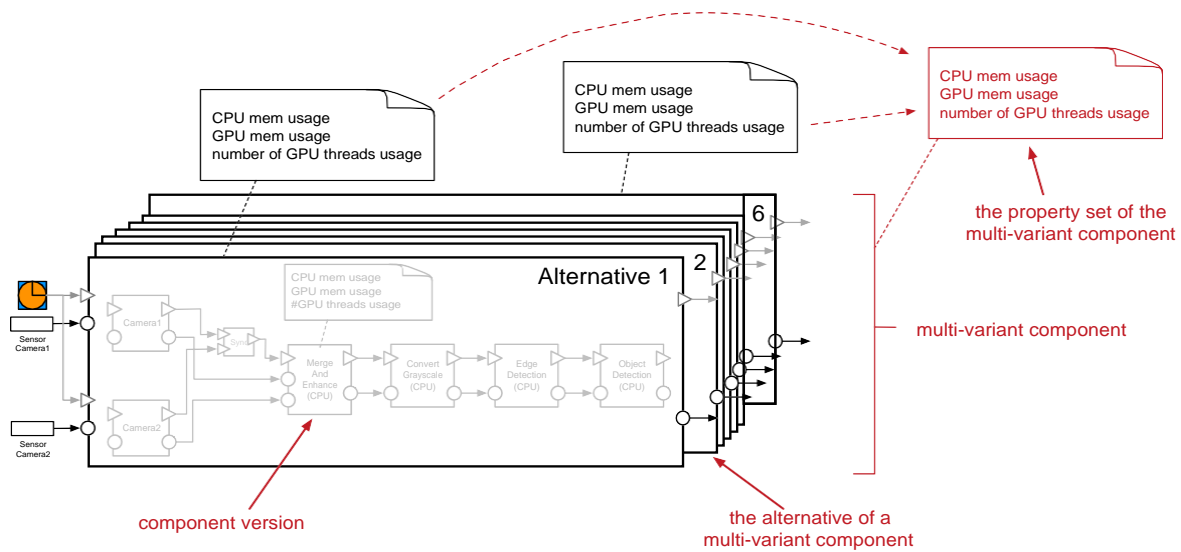

**Figure 3.** A multi-variant vision component.

In general, using a simplified version of the system architecture decreases the amount of information forwarded to the allocator, and hence, the allocation complexity. The simplified version is obtained by applying a high-level layer on top of the detailed architecture layer. Instead of describing each component by a set of requirements as it is done in the detailed layer, connected components are grouped into units with the same functionality, that behave as regular components, and are described by a condensed number of properties as it is achieved in the high-lever layer. Furthermore, some desired allocation constraints may be automatically taken care of and provided as a by-product of our method. For example, instead of specifically requesting that connected CPU–GPU components to be allocated together onto the same CPU–GPU chip units (e.g., to improve the system performance), our method implicitly introduces this request by compressing the components into single (component-like) units.

We construct our solution around an existing allocator. The (mathematically defined) model of the allocator is based on constraints and optimization goals. The constraints assure that the allocator does not use more than existing hardware resources such as the memory required by all the components placed on a platform does not exceed the available physical memory. The optimization goals allow the user to determine essential features of the system such as performance. The actual allocation is done by using a constraint solver, the details of which we have described in a previous paper [3].

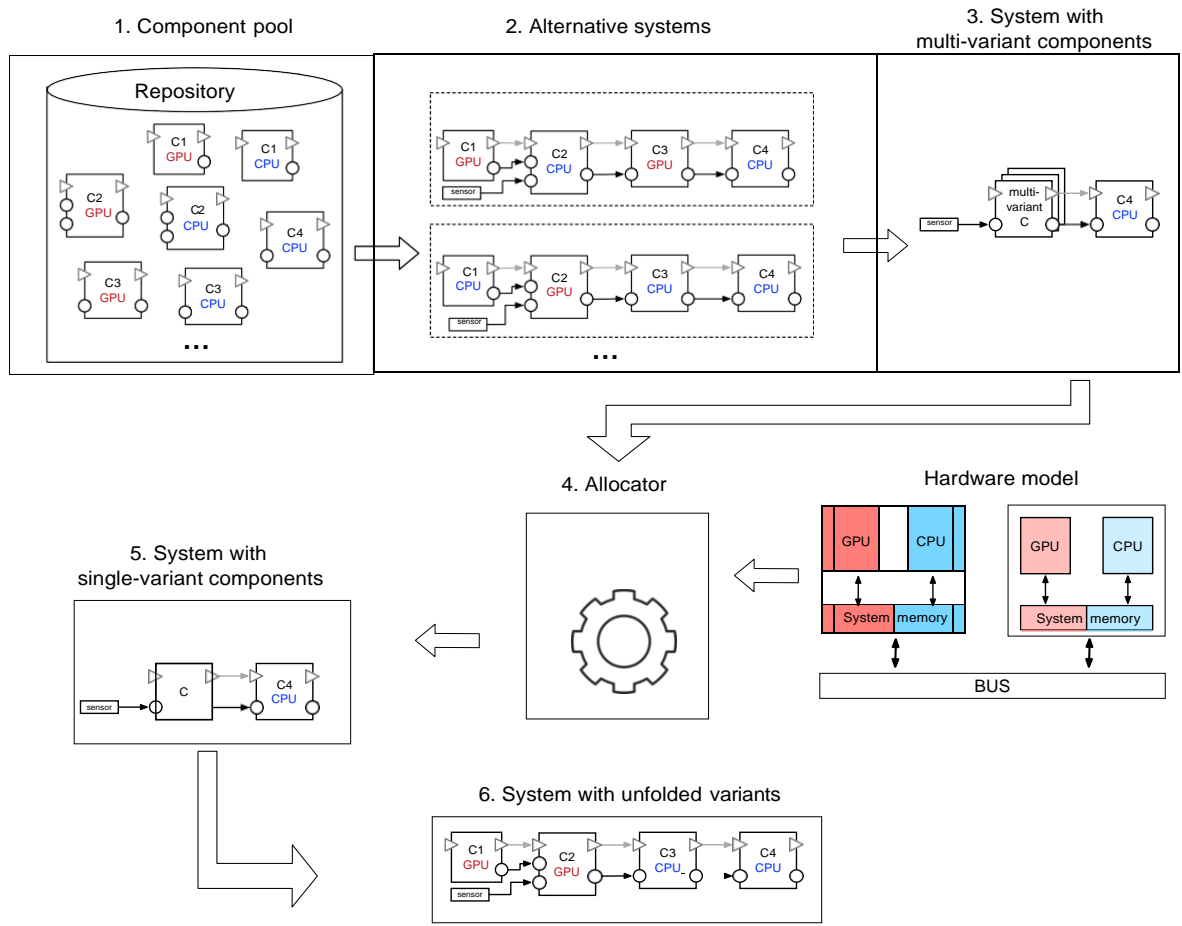

**Figure 4.** The overview description of our solution.

## 4. Evaluation

In this section, we analyze our solution in two parts. In the first part, we look into the feasibility aspect where the proposed solution is implemented using an existing system. The second part analyzes the scalability aspect of the solution.

*4.1. Feasibility*

To analyze the feasibility of our solution, we apply our allocation method on the underwater robot, partially introduced in Section 2. The robot contains two connected embedded controller boards: (i) a System-on-Chip board that contains a CPU–GPU chipset, and (ii) a regular board with a one-core ARM CPU. The boards communicate via a Controller Area Network (CAN) bus, and are connected to various sensors (e.g., cameras, pressure sensor) and actuators (e.g., thrusters). We characterize each board by a set of properties as follows:

- *availMem* represents the available memory of the board, and is measured in megabyte (MB).
- *availCpu* represents the available load of the CPU and its value is compared to a reference unit (e.g., 1 Cpu load unit is a particular amount of work over a period of time).

- *availGpu* represents capacity of a GPU. As a metric we use the amount of threads a GPU has. Although it is not an accurate description, we use this measurement unit to characterize, at a high-level, the GPU power.

A high-level model of the underwater robot is presented in the right-hand side of Figure 5. We notice that for the board that has only a (one-core) CPU, the GPU-related property (i.e., *availGpu*) is set to zero.

The robot is equipped with two vision systems, the front system using two from cameras and the bottom system using a single bottom camera. For both vision systems we constructed different CPU- and GPU-based components as follows. The front vision functionality merges two RGB frames, converts the merged frame into the grayscale format, applies an edge-detection filter and detects the objects from the frame (for more details see Section 2). To construct the functionality, we developed: (i) six CPU-based components (i.e., Camera1, Camera2, MergeAndEnhance CPU, ConvertGrayscale CPU, EdgeDetection CPU and ObjectDetection GPU), and (ii) four GPU-based components (i.e., MergeAndEnhance GPU, ConvertGrayscale GPU, EdgeDetection GPU and ObjectDetection GPU).

The bottom vision has a similar functionality. Due to the fact that there is only one camera, there is no need to merge frames. Therefore, to construct the bottom vision functionality, we reused four of the components developed for the front vision. In this case, there are: (i) four CPU-based components (i.e., Camera1, ConvertGrayscale CPU, EdgeDetection CPU and ObjectDetection GPU), and (ii) three GPU-based components (i.e., ConvertGrayscale GPU, EdgeDetection GPU and ObjectDetection GPU).

Besides the vision systems, five more CPU-based components are also needed. The VisionManager component takes vision decisions based on the data received from the front and bottom vision systems. DecisionCenter is the brain of the system which controls, based on the information received from VisionManager, the system settings (e.g., water pressure) and selects between the robot missions (e.g., find red buoys). The robot thrusters are managed by the MovementNavigation component that maneuvers the underwater robot using the data received from the DecisionCenter component.

Each component is characterized by the following properties:

- *reqMem* characterizes the memory usage requirement of a component and is measured in MB.
- *reqCpu* presents the CPU usage requirement of a component.
- *reqGpu* describes the component GPU usage requirement and is measured in number of threads.
- *Exec* is related to the performance of the component and describes the execution time expressed in milliseconds.

These components, constructed by the component developer, are placed into a Component repository which is illustrated in the upper part of Figure 5. The system developer uses the available components and constructs the system architecture. The (front and bottom) vision systems have multiple alternatives as illustrated in Figure 5. Each alternative has a distinct set of properties and when all alternatives are combined into a single multi-variant component, the properties are described as a sequence of values. Each value of the sequence represents the resource usage of the corresponding variant. For example, the FrontVision multi-variant component requires, for its first alternative (i.e., all CPU-based components), 6 MB of memory, 0.6 CPU load, 0 GPU threads and has an execution time of 22 ms.

Our solution is constructed around an existing allocator. In the following paragraphs, we introduce the formal model of the allocator. It contains three parts, i.e., the input, the constraints and the optimization function, as follows.

1. The input part describes the software components and the platform. Let be $C$ a set of $n$ software components, and four functions $reqMem : C \rightarrow \mathbb{Q}^+$, $reqCpu : C \rightarrow \mathbb{Q}^+$, $reqGpu : C \rightarrow \mathbb{Q}^+$ and $Exec : C \rightarrow \mathbb{Q}^+$, where:

$$
\begin{aligned}
reqMem(c) &= \text{the required memory of component } c \\
reqCpu(c) &= \text{the CPU workload required by component } c \\
reqGpu(c) &= \text{the GPU threads required by component } c \\
Exec(c) &= \text{the execution time of } c
\end{aligned}
$$

The platform is characterized by a set $H$ of $k$ computation nodes (i.e., either CPU or CPU–GPU based), and three functions $useMem : H \rightarrow \mathbb{Q}^+$, $useCpu : H \rightarrow \mathbb{Q}^+$ and $useGpu : H \rightarrow \mathbb{Q}^+$, where:

$$
\begin{aligned}
useMem(h) &= \text{the usable memory on node } h \\
useCpu(h) &= \text{the CPU capacity on node } h \\
useGpu(h) &= \text{the available number of GPU threads on node } h
\end{aligned}
$$

2. The constraints are defined in order to ensure a feasible allocation. Given the allocation function $allocation : C \rightarrow H$, we define the following constraints:

   - The usable memory of a node $h$ should not be exceeded by the summed required memory of components placed on $h$.

   $$\forall h \in H \left( \sum_{c \in C | h \in allocation(c)=h} reqMem(c) \leq useMem(h) \right)$$

   - The usable CPU workload of a node $h$ should not be exceeded by the summed required workload of components placed on $h$.

   $$\forall h \in H \left( \sum_{c \in C | h \in allocation(c)=h} reqCpu(c) \leq useCpu(h) \right)$$

   - The total amount of GPU threads of a node $h$ should not be exceeded by the summed number of threads required by the components placed on $h$.

   $$\forall h \in H \left( \sum_{c \in C | h \in allocation(c)=h} reqGpu(c) \leq useGpu(h) \right)$$

3. The optimization function:

$$P(allocation) = \sum_{c \in C | h \in allocation(c)=h} Exec(c)$$

provides the best performance of the allocation:

$$\text{minimize } (P)$$

The system properties and the constraints and optimization goal are fed to the allocator that computes allocation schemes. The allocation model alongside with all its required information (i.e., system properties, constraints and optimization goals) are translated into the IBM-CPLEX solver. The advantage of employing a mathematical solver is that the computed solution is optimal.

The front vision is considered the main vision of the robot, while the bottom vision is seen as a secondary system being activated when e.g., the main vision does not detect any objects. Therefore, the front vision priority is higher in accessing the GPU resources. The allocator computes allocation schemes as presented in Figure 5, where both of the vision systems are allocated onto the $H_1$, and the rest of the system is allocated onto $H_2$. With a higher priority, the front vision accesses more GPU resources (i.e., threads) than the bottom system.

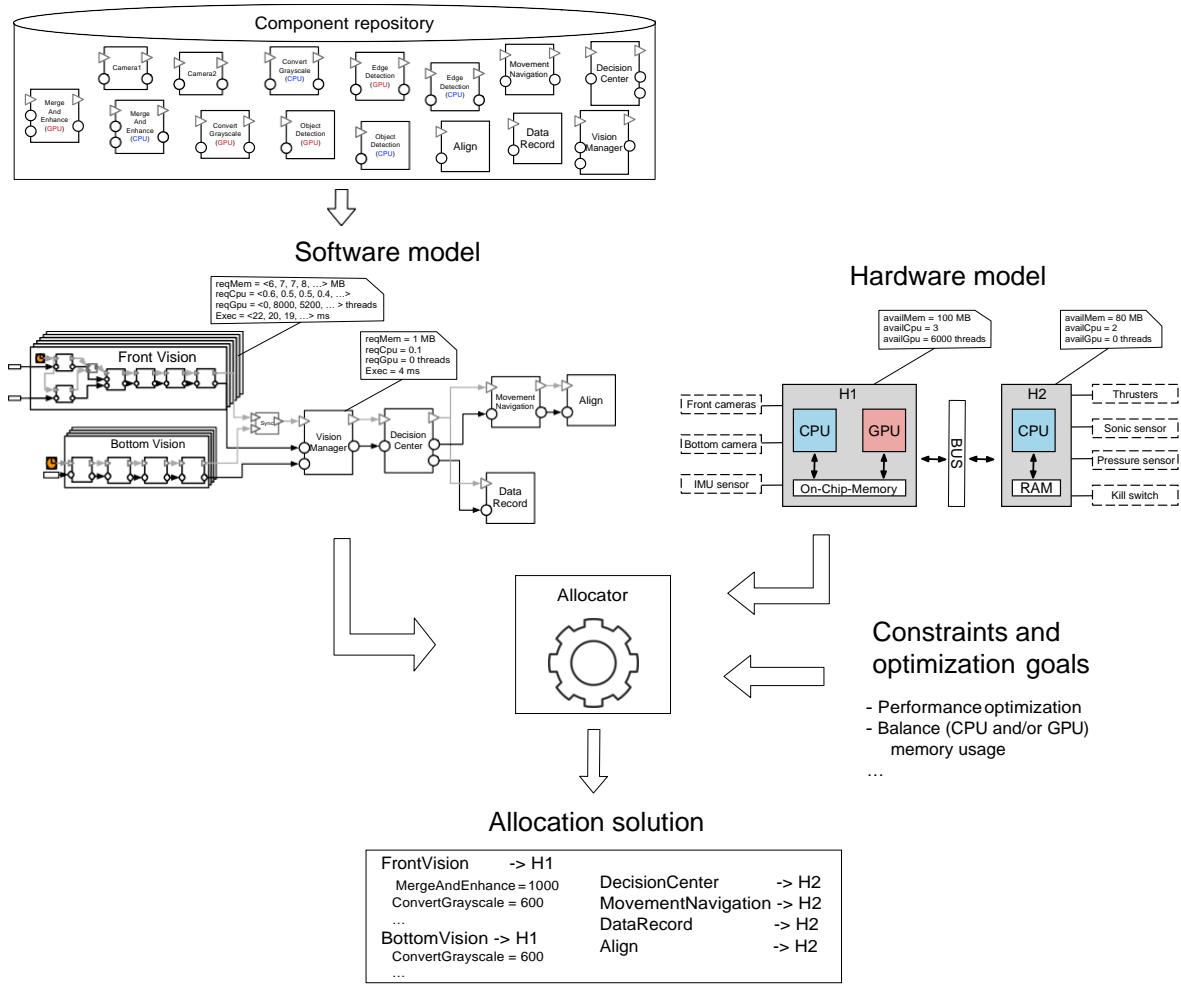

**Figure 5.** The allocation process of the underwater robot.

The optimal result computed by the CPLEX solver is (partially) described in Figure 6 where the selected front vision alternative contains four GPU-based components and the bottom vision alternative contains only one GPU-based component. The system description corresponds to the detailed architecture view, where the selected single-variant components are unfolded.

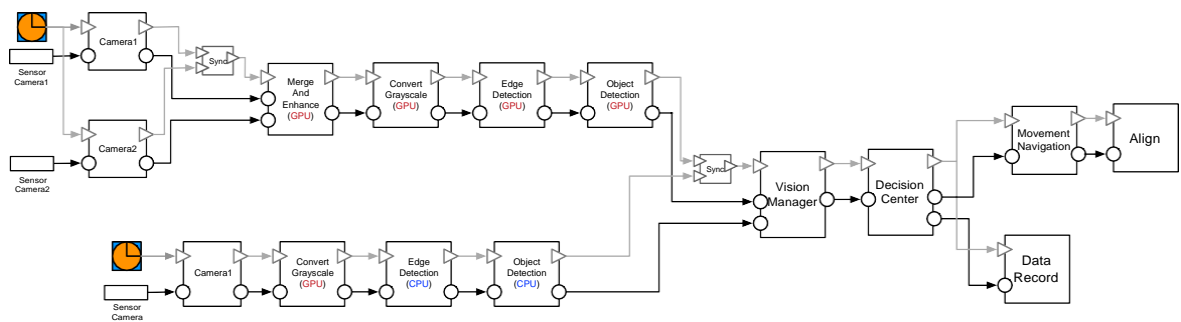

**Figure 6.** The allocation scheme result with unfolded variants.

The computed solution is a feasible system considering the available hardware resources and the configured optimization goals.

*4.2. Scalability*

For the scalability analysis we designed a test system composed of *n* + 1 chained components. Each of the first *n* components has two versions with the same functionality, i.e., one version with CPU-based functionality and the other with GPU-based functionality. For this part of the evaluation, we compare two versions of the system, where one version (referred as the naïve system) contains a chain of *n* +1 components, and the other version contains a multi-variant component.

The multi-variant component is constructed from the different alternatives that result from composing different component versions. For simplicity, we consider only a two-variant component, where one alternative contains all *n* components with CPU-based functionality, while the other alternative contains all *n* components with GPU-based functionality, as illustrated by Figure 7. Each individual component is characterized by memory usage, CPU usage, GPU thread usage and execution time. The multi-variant component is characterized by a set of properties derived from the enclosed individual component properties, where each property is described as a sequence of the values (i.e., one for the CPU- and the other for the GPU-based components). For example, one value of the memory sequence property that characterizes the CPU-based component variant, is computed by summing the memory usage requirements of all its *n* CPU-based components.

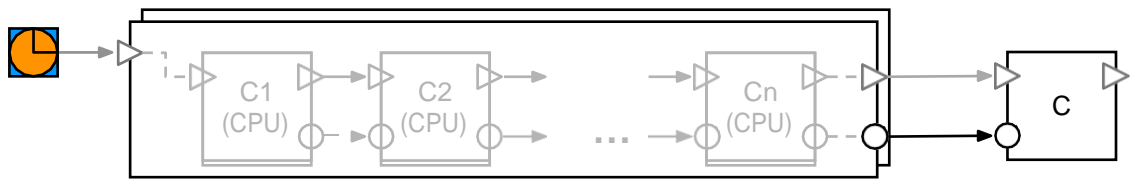

**Figure 7.** A system composed of chained components.

To calculate the allocation computation time, we have constructed three systems; the first system contains 31 components (i.e., *n* = 30), the second system contains 41 components (i.e., *n* = 40) and the last system has 51 components (i.e., *n* = 50). We randomly provided values for component properties (i.e., for each CPU- and GPU-based version). For instance, the component memory usage is randomly assigned a value between 1 and 100. Regarding the hardware, we assume that we have six connected boards, where only three of them have GPUs. Similarly, we characterized the boards resources in a random manner. For example, the available memory of a board is randomly assign a value between 100 and 2500.

Using the implemented CPLEX allocator from the previous part, we compute allocation schemes for our test systems. As optimization goal, we set the allocator to provide the best performance (i.e., execution time).

The scalability results are presented by Table 1. Using a machine with a 2.6 GHz i7 CPU and 16 GB of memory, we ran the allocation 1000 times for three systems, i.e., a naïve system that contains CPU-based/GPU-based components, and a system that contains the two-variant component. The results show that the allocator uses less time to compute results for the system with the two-variant component. In other words, for the CPU- and GPU-based systems, where there are *n* + 1 components and each has its own set of properties, the allocator analyzes a higher number of properties than for the two-variant system where the two-variant component has one set of properties. Furthermore, the computation time for the CPU-based system is relatively the same as for the GPU-based system due to the fact that the analyzed number of properties are the same for the two systems. By proposing the two-layer allocation design, we show in this part of the evaluation, how the scalability of an allocator for heterogenous CPU–GPU systems is improved.

Table 1. The time used to compute allocation schemes.

| | Average Allocation Time (ms) | | |
|---|---|---|---|
| $n$ | Naïve * | | Two-Variant ** |
| | CPU-Based | GPU-Based | |
| 30 | 18.2 | 18.3 | 15.4 |
| 40 | 29.1 | 29.0 | 24.6 |
| 50 | 49.5 | 49.7 | 39.8 |

\* A system with (CPU-based/GPU-based) $n + 1$ components; ** A system with a two-variant component; GPU, graphics processing units; CPU, central processing unit.

**5. Related work**

We introduce in [3] the initial idea of the two-layer allocation method. We extended the initial work by describing the solution using an existing component model and presenting the overview using existing components. Furthermore, in the current paper we introduced an existing system and applied the solution on it to analyze the feasibility aspects. Moreover, the evaluation section analyzes the scalability aspects, which were not covered by the previous work.

Software-to-hardware allocation and optimization of the allocation mechanism have been the topic of many research works in the literature. A systematic literature review on the software architecture optimization methods is provided in [10]. The authors in this work analyzed 188 papers and identified 30 papers related to the optimization of component-based systems. Of this set of papers, only 13% (i.e., 4 papers) use exact optimization algorithms (vs. approximate algorithms). While exact optimization algorithms can provide optimal solutions, their applications poses several challenges when it comes to adopting them, such as difficulty of formally defining the allocation model, search-space, and the usually non-linearity of the object functions (and thus being computationally expensive). The approach we presented in this paper, enabled us to formally define our optimization model which in turn allows to use exact optimization algorithms and methods. Moreover, one of the important characteristics of our approach is that, as we demonstrated, it simplifies the search-space and therefore complexity for allocation optimization.

Consideration of quality attributes and satisfaction of non-functional requirements play an important role in designing embedded systems due to resource constraints of these systems. While our proposed approach can address different quality attributes (such as memory, processing capacity and number of threads, etc.) and is generic in this regard, there are some research in the literature that target specific quality attributes. For instance, in [11], a detailed optimization model and framework for energy consumption in component-based distributed systems in Java is provided. The main goal in this work has been to help system architects make informed decisions such that the energy consumption is reduced in a designed system. An interesting aspect discussed in this work is the energy consumption of the communication of components that reside in different Java Virtual Machines, on the same host. From this perspective, the communication aspect is implicitly address in our optimization model where connected CPU- and GPU-based components are tried to be allocated on the same node. Furthermore, in our optimization model, the energy is treated as any other property in a generic fashion. [12] is another example of works that address energy usage in heterogeneous multiprocessor embedded systems. In this work, an optimization model using integer linear programming is introduced that minimizes the system energy usage when the end-to-end time constraints are given. Moreover, the CPLEX solver is used to compute allocation solutions. It is shown that, for a system with more than 30 components, the solver computes solutions in up to couple of minutes. The solution introduced in our work aims at decreasing the allocator burden and we show that, for a system with 31 components, the allocation computation time is reduced.

Wang et al. in [13] introduce a method to allocate the software components in a design model to a given platform while meeting multiple platform resource constraints. In the method, different types of resources are considered and weights are used to define the importance of each in the allocation process. The components that require more resources get higher priority getting allocated first. In contract, in our approach all components have same and equal allocation priority with respect to their resource requirements. Also using the flexible component concept, we increase the flexibility of the allocation regarding the component resource requirements. Weight parameters are used in our approach to define the importance of properties in the allocation process.

For the systems in the automotive domain [14] proposes optimization of software allocation and deployment to hardware nodes (i.e., ECUs) as a as a bi-objective problem using an evolutionary algorithm. It considers the reliability of data communications between components as one and the communications overhead as a second objective.

## 6. Conclusions

In this paper, we proposed a method to reduce the burden of the allocator and ease the software-to-hardware allocation in the design of component-based embedded systems. Our solution works by introducing a two-layer architecture design for heterogeneous CPU–GPU embedded systems, where detailed component information is abstracted using the two layers to ease allocation decisions. The work is independent of the employed allocator, i.e., how the allocator itself is implemented and based on which solver.

To show the feasibility of the approach, we applied it on a underwater robot which we participated in its development. Additionally, to also demonstrate and evaluate the scalability of the approach, we analyzed it on three test systems consisting of 31, 41, and 51 components respectively. We compared the average allocation time for two versions of each of these test systems: (i) containing all components, (ii) a two-variant component model of the system, based on the two-layer allocation concept. The results show that the allocator does its computations faster and requires less time for the two-variant component version. Although CPLEX solver was used in this work, the proposed solution can be implemented in any mixed-integer non-linear solver. However, the usage of a different solver may influence the allocation computation time.

In terms of quality attributes and non-functional properties, the proposed two-layer allocation solution is generic and be be applied in allocation optimization based on any set of properties. From this perspective, it is property-agnostic. Deriving variant properties (i.e., aggregated from its constituting components), however, can be less trivial for certain non-functional properties such as energy. In our case study in this paper, we characterized the system through simple properties such as static memory or GPU thread usage. Then to derive the multi-variant properties, we simply used addition operation for these properties. As a future direction, we plan to investigate how energy usage can be derived for variants, and thus, enable its inclusion and evaluation as part of our proposed solution. Another extension of this work is to extend the scope of heterogeneity of the approach to include other processing units such as DSPs and FPGAs as well.

4. AUTOSAR—Technical Overview. Available online: http://www.autosar.org (accessed on 27 July 2018).
5. John, K.H.; Tiegelkamp, M. *IEC 61131-3: Programming Industrial Automation Systems: Concepts and Programming Languages, Requirements for Programming Systems, Decision-Making Aids*; Springer Science & Business Media: Berlin, Germany, 2010.
6. Torngren, M.; Chen, D.; Crnkovic, I. Component-based vs. model-based development: A comparison in the context of vehicular embedded systems. In Proceedings of the 31st EUROMICRO Conference on Software Engineering and Advanced Applications, Porto, Portugal, 30 August–3 September 2005; pp. 432–440. [CrossRef]
7. Saadatmand, M.; Leveque, T. Modeling Security Aspects in Distributed Real-Time Component-Based Embedded Systems. In Proceedings of the 2012 Ninth International Conference on Information Technology—New Generations, Las Vegas, NV, USA, 16–18 April 2012; pp. 437–444. [CrossRef]
8. Ahlberg, C.; Asplund, L.; Campeanu, G.; Ciccozzi, F.; Ekstrand, F.; Ekstrom, M.; Feljan, J.; Gustavsson, A.; Sentilles, S.; Svogor, I.; et al. The Black Pearl: An autonomous underwater vehicle. In Proceedings of the AUVSI Foundation and ONR's 16th International RoboSub Competition, San Diego, CA, USA, 22–28 July 2013.
9. Hänninen, K.; Mäki-Turja, J.; Nolin, M.; Lindberg, M.; Lundbäck, J.; Lundbäck, K.L. The Rubus component model for resource constrained real-time systems. In Proceedings of the International Symposium on Industrial Embedded Systems, Le Grande Motte, France, 11–13 June 2008; pp. 177–183.
10. Aleti, A.; Buhnova, B.; Grunske, L.; Koziolek, A.; Meedeniya, I. Software Architecture Optimization Methods: A Systematic Literature Review. *IEEE Trans. Softw. Eng.* **2013**, *39*, 658–683. [CrossRef]
11. Seo, C.; Malek, S.; Medvidovic, N. An Energy Consumption Framework for Distributed Java-based Systems. In Proceedings of the Twenty-Second IEEE/ACM International Conference on Automated Software Engineering, Atlanta, GA, USA, 5–9 November 2007; ACM: New York, NY, USA, 2007; pp. 421–424. [CrossRef]
12. Goraczko, M.; Liu, J.; Lymberopoulos, D.; Matic, S.; Priyantha, B.; Zhao, F. Energy-optimal software partitioning in heterogeneous multiprocessor embedded systems. In Proceedings of the 2008 45th ACM/IEEE Design Automation Conference, Anaheim, CA, USA, 8–13 June 2008; pp. 191–196. [CrossRef]
13. Wang, S.; Merrick, J.R.; Shin, K.G. Component allocation with multiple resource constraints for large embedded real-time software design. In Proceedings of the 10th IEEE Real-Time and Embedded Technology and Applications Symposium, Toronto, ON, Canada, 25–28 May 2004; pp. 219–226. [CrossRef]
14. Moser, I.; Mostaghim, S. The automotive deployment problem: A practical application for constrained multiobjective evolutionary optimisation. In Proceedings of the IEEE Congress on Evolutionary Computation, Barcelona, Spain, 18–23 July 2010; pp. 1–8. [CrossRef]